# Characterization of the new scintillator — $Cs_2LiYCl_6:Ce^{3+}$ *

Li Kui-Nian (李奎念)[1;1)]   Zhang Xian-Peng (张显鹏)[1, 3]   Gui Qiang (桂强)[2)]   Jin Peng (金鹏)[1)]   Tian Geng (田耕)[1)]

1 Northwest Institute of Nuclear Technology, Xi'an 710024, China

2 Beijing Glass Research Institute, Beijing 101111, China

3 School of Nuclear Science and Technology, Xi'an Jiaotong University, Xi'an 710049, China

Abstract:  The first domestic inorganic scintillator $Cs_2LiYCl_6:Ce^{3+}$（CLYC）was grown in Beijing Glass Research Institute using the vertical Bridgman method. In this work, we evaluated the performance of this new CLYC crystal in terms of gamma-ray energy resolution and pulse shape discrimination (PSD) capability between neutrons and gamma-rays. The decay times occurred through different scintillation mechanisms were achieved by fitting decay functions to the neutron and gamma-ray waveform structures. We found an energy resolution of ~4.5% for 662-keV gamma-rays and efficient neutron/gamma PSD with FoM ~2.6. Under gamma-ray excitation, there is ultrafast scintillation mechanism in CLYC, with a decay time about 2 ns, whereas there is no evidence of ultrafast decay under thermal neutron excitation. This work contributes to promote domestic development of CLYC.

PACS：29.40.-n, 29.40.Mc, 81.10.Fq

* Project supported by the National Natural Science Foundation of China (Grant No. 11575145)

1) Corresponding author. E-mail: lk.nian@foxmail.com

## 1 Introduction

$Cs_2LiYCl_6:Ce^{3+}$ (CLYC[1] is a promising new inorganic scintillator. It was first discovered as a scintillator in 1999 by researchers from Delft University of Technology[2]. They reported that CLYC had the capability for thermal neutron detection based on the $^6Li(n,\alpha)$ reaction and made further investigations on CLYC in next few years[3-5]. Van Loef revealed the scintillation mechanism of CLYC, known as Core-to-Valence-Luminescence (CVL), in 2005[6]. The CVL has an ultrafast decay time (~ns) and appears only under gamma-ray radiation[7]. CVL enabled CLYC to discriminate between neutron (alpha) particles and gamma-rays with pulse shape discrimination (PSD) methods[8]. The capability for PSD of CLYC drew great attentions from the nuclear science community[9-14]. Under gamma-rays irradiation, its light yield is about 20000 photons/MeV[15], and it exhibits good energy resolution of ~4% FWHM @662keV due to excellent linearity[16]. Under thermal neutrons irradiation, the light yield is about 70500 photons per neutron. The thermal neutron peak resolution is typically 2.5~3.1%[9].

Radiation Monitoring Devices Inc. grew its first CLYC crystal in 2003. Since then, a number of CLYC crystals with different sizes and features have been grown with the development of crystal growth technology[17-19]. In 2012, CLYC became available through commercial approaches, but the price was still high. To establish the domestic growth technology of CLYC, Beijing Glass Research Institute (BGRI) grew a promising CLYC crystal in 2016. To our knowledge, it is the very first domestic CLYC crystal. To evaluate the performance of the new CLYC crystal, we focused on two important properties of it: gamma-ray energy resolution and PSD capability. Moreover, we achieved the decay times of different scintillation mechanism by fitting decay functions to the neutron and gamma-ray waveform structures.

## 2 Experimental

### 2.1 Crystal growth

CLYC crystal was grown using the vertical Bridgman method in BGRI. In this method, stoichiometric amounts of CsCl, LiCl, YCl$_3$ and CeCl$_3$ were mixed in a quartz crucible. Natural LiCl that contains about 7% of the $^6$Li isotope was used. The dopant (Ce$^{3+}$) concentration was 0.5mol% as a standard doping level. After mixing powders in the quartz crucible, vacuum was pumped to $10^{-5}$ Pa to avoid oxidization. The quartz crucible was sealed, before fusing the mixed powders in a stove at a temperature above 640 ℃. Along with lowering the quartz crucible steadily with a velocity of 2 millimeter per hour, the CLYC crystal ingot was grown.

After cutting and polishing, the CLYC crystal was packed in an aluminum casing with a reflective inner coating and sealed with a quartz window at one end. CLYC has density of 3.31g/cm$^3$ and its peak-emission wavelength is about 390nm, which is well matched to many commercially photomultiplier tubes. Fig.1 shows the picture of the CLYC crystal with 25.4 mm diameter and 15 mm thickness.

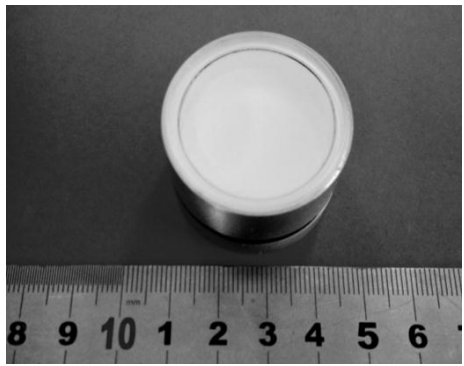

Fig.1. Photograph of CLYC crystal with 25.4mm diameter and 15mm thickness grown in BGRI. The crystal is packaged in an aluminum enclosure.

2.2 Data acquisition

The CLYC crystal was coupled with a 2 inch diameter 9815B blue-green sensitive bialkali photomultiplier (PMT) with optical grease. 9815B PMT's quantum efficiency is ~30% at peak-emission wavelength (390nm) of CLYC. To estimate gamma-ray energy resolution of CLYC crystal, a $^{137}$Cs source was used as gamma-ray source. The signals from PMT anode with different heights were analyzed in a multi-channel spectroscopy by using a shaping time of 4 microseconds.

An Am-Be source which emits approximately $1\times10^4$ neutrons per second was used as neutron and gamma-ray source to estimate the PSD capability of CLYC. The schematic diagram of the experimental setup is shown in Fig.2. A Tektronix DPO7104 oscilloscope was set to FastFrame mode (1 G samples per second and 2 us recording length) to record the PMT anode signals from both neutrons and gamma-rays with a 50 Ohm terminator. The CLYC crystal coupled with 9815B PMT was placed in a polyethylene tube. A polyethylene block with 2.5cm thickness and a lead block with 2.5cm thickness were placed between the CLYC crystal and Am-Be source. Approximate 5000 signals were recorded, and analyzed off-line with PSD method (see section 3.2).

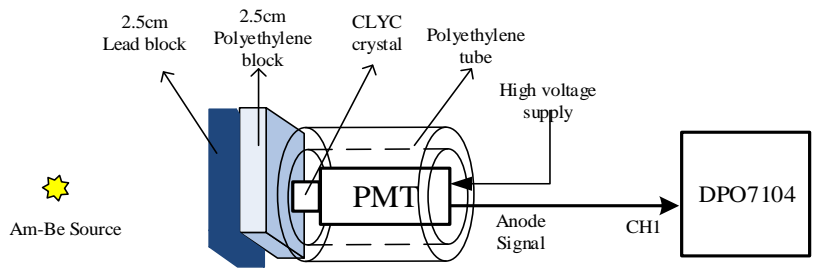

Fig.2. Experimental setup in Am-Be source. The CLYC crystal was coupled with a 9815B PMT which connected directly to a digital oscilloscope (Tektronix DPO7104)

## 3 Results and analysis

### 3.1 Gamma-ray energy resolution

Fig. 3 shows the energy spectra measured with CLYC crystal under irradiation from $^{137}$Cs gamma-ray (662keV) source. The position of the 662keV energy peak was used to scale the horizontal axis in energy units (keV). The CLYC energy resolution at the 662keV full energy peak is ~4.5% (full width at half maximum, FWHM) , which is better than those commonly used scintillators in gamma detection, such as NaI(Tl) and CsI(Tl), having 6.5% and 7-9% energy resolution @662keV respectively.

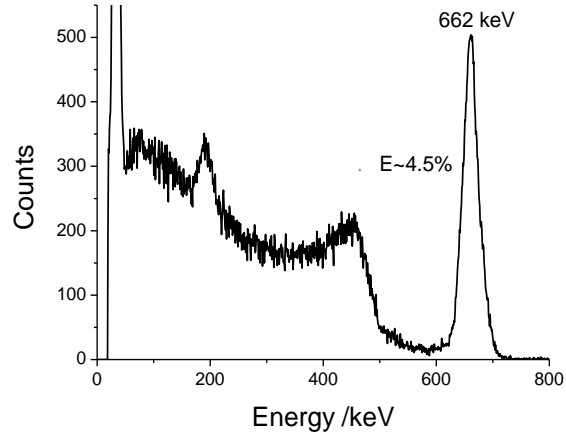

Fig.3.  $^{137}$Cs energy spectra measured with CLYC crystal. An energy resolution of ~4.5% FWHM @662keV was acheived.

3.2  Neutron and gamma-ray discrimination

PSD was applied to separate gamma-ray from neutron waveforms based on the traditional charge integration method. In this method, two integration windows of the signal pulses, named as prompt integration window and delays integration window individually, were selected and integrated. The prompt window includes the sharp rise and peak portion, and the delayed window includes the decaying portion of the signals. Generally, there are infinite combinations of possible integration window widths. It is essential for achieving the best possible PSD by optimizing window widths. The automated waveform analysis algorithm can optimize window width through the following procedure:

(1) Aligning the waveforms to a common zero. The point where the amplitude was 20% of the peak value was chosen as the common zero.

(2) Charge integration according to two integration windows. The width of the prompt integration window ranged from 5 ns to 300 ns by 5 ns. The delayed integration window started right after the end point of the prompt window and its width ranged from 10 ns to 1300 ns by 10 ns.

(3) Calculating the PSD ratio. For each unique combination of the prompt and delayed integration windows, PSD ratio was calculated for each waveform based on charge integrals $Q_{prompt}$ and $Q_{delayed}$. The PSD ratio was defined as $Q_{delayed}/Q_{prompt}$.

(4) Calculating the Figure-of-Merit (FoM). FoM was utilized to evaluate the performance of the neutron and gamma-ray discrimination. FoM was defined as

$$\text{FoM} = \frac{D}{n_{FWHM} + \gamma_{FWHM}} \qquad (1)$$

Where $D$ is the distance between the centroids of the neutron and the gamma peak in the PSD ratio histogram, and $n_{FWHM}$ and $\gamma_{FWHM}$ are the corresponding FWHMs. To obtain the centroid positions and FWHMs, Gaussian fits were performed to the neutron and gamma-ray peaks in the histogram.

The result of the optimization for PSD is shown in Fig.4. This figure shows that FoM is more sensitive to the prompt window width compared with the delayed window width. With increasing prompt window width, FoM shows a fast and short increase, followed by a sharp, long decrease. Change of delayed window width has a minimal effect on the FoM. For the CLYC crystal in this study, the best FoM we found was 2.6, in a prompt window width of 20 ns and delayed window width of 900ns.

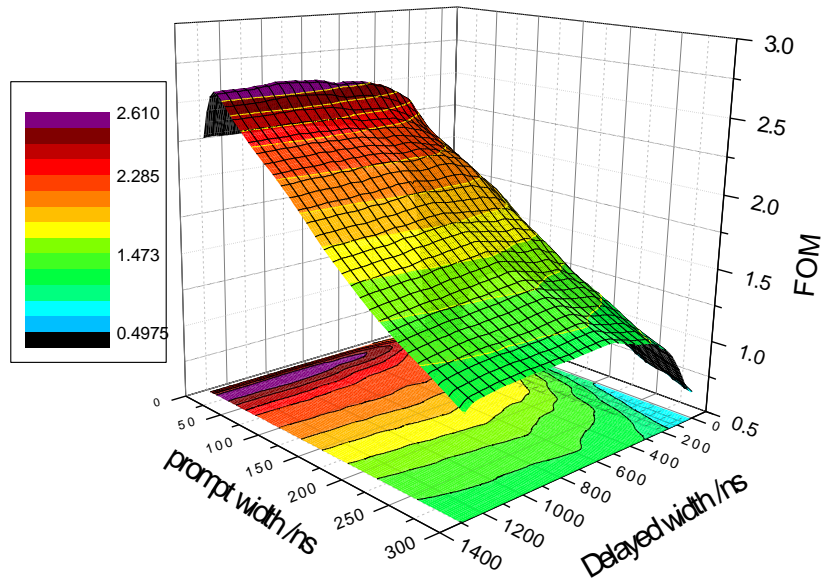

Fig.4. (color online) FoM values under different prompt and delayed integration windows combinations. The prompt width ranges from 5 ns to 300 ns by 5 ns. The delayed width range from 10 ns to 1300 ns by 10 ns.

With the optimal prompt and delayed integration windows, a two-dimensional PSD scatter plot was generated for the neutron and gamma-ray events from an Am-Be source (see Fig.5). In Fig.5, gamma-ray events and neutron events show us two disparate distribution patterns. Gamma-ray events have lower PSD ratio across the wide energy range while neutron events are isolated with higher PSD ratio. The neutron events can be divided into three groups based on different reaction: fast neutrons ($^{35}$Cl(n,p) reaction), thermal neutrons ($^{6}$Li(n,α) reaction) and fast neutrons ($^{6}$Li(n,α) reaction), as indicated in Fig.5. Fig.6 (dash line) shows the one-dimensional projection of neutron and gamma-ray events corresponding to the Y-axis in Fig.5. A Gaussian function was fitted to the neutron and gamma-ray peaks in Fig.6 (solid line). The centroids and FWHMs obtained from the Gaussian fits were used to calculate FoM value.

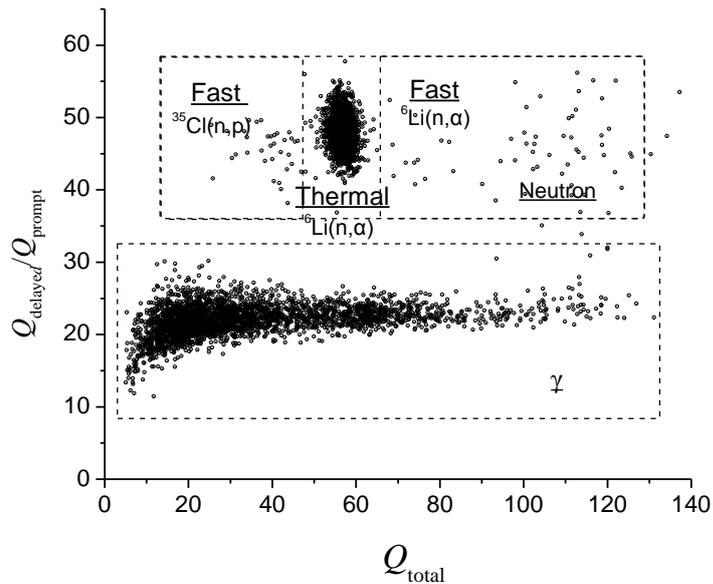

Fig.5. Two-dimensional PSD scatter plot based on the data collected under irradiation from an Am-Be source. X-axis is the PSD ratio ($Q_{delayed}/Q_{prompt}$) and Y-axis is the total integral $Q_{total}$ corresponding to the energy of events. Gamma-ray events (on the bottom) and neutron events (on the top) are separated clearly. Fast neutrons ($^{35}$Cl), thermal neutrons ($^{6}$Li) and fast neutrons ($^{6}$Li) are indicated.

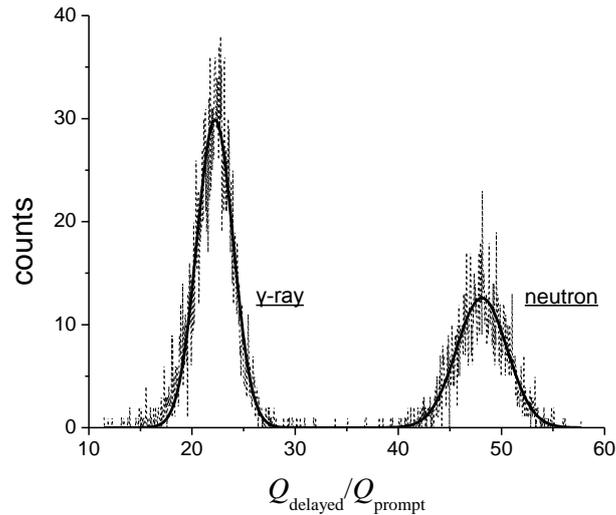

Fig.6. One-dimensional projection of neutron and gamma-ray events to the Y-axis in Fig.5. A Gaussian function was fitted to the neutron and gamma-ray peaks (solid line).

3.3 Pulse shape analyses

According to the PSD optimal results, neutron signals and gamma-ray signals were averaged respectively to produce the standard-pulses. These standard-pulses contribute to eliminate noise fluctuations in individual signals and reveal the details of the prompt and decayed pulse shape. Neutron standard-pulse and gamma-ray standard-pulse were inverted and normalized to their peaks, as shown in Fig. 7. Significantly different characteristics can be found between neutron pulse and gamma-ray pulse due to different scintillation mechanisms.

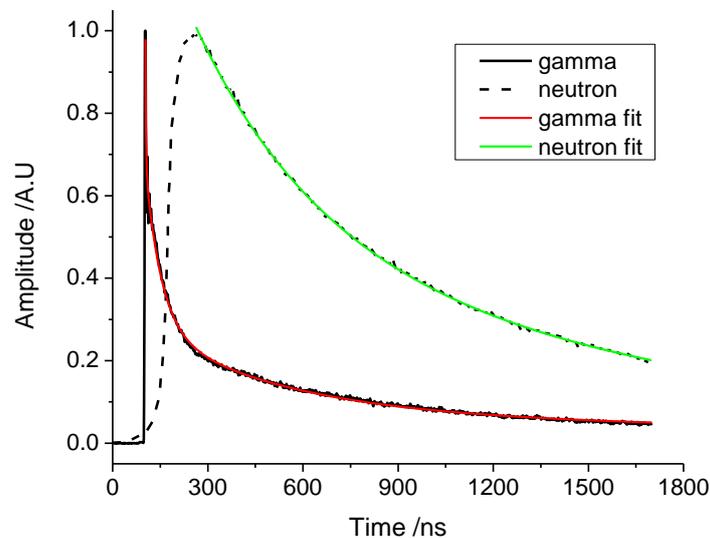

Fig. 7. (color online) Overlay of neutron and gamma-ray standard-pulse. Exponential decay

curves fitting of the standard pulses are shown. The red solid line is the gamma-ray standard-pulse exponential fitting curve and the green solid line is the neutron standard-pulse exponential fitting curve.

Generally, for Ce3+-doped CLYC crystal, up to four mechanisms contribute to the scintillation[10,16]: $Ce^{3+}$ (direct electron-hole capture by $Ce^{3+}$), $V_k$ (binary $V_k$-electron diffusion), STE (self-trapped exciton emission), and CVL (core-to-valence luminescence).

The first scintillation mechanism $Ce^{3+}$ may occur when an electron that is created in the valence band is excited into the conduction band of the crystal by the incident gamma-ray. The formed free electron-hole pair may be captured directly by a $Ce^{3+}$ ion leading to 4f-5d excitation and 5d-4f emission which produce a prompt emission with very fast rise time. If the hole gets trapped by two anions, a molecular complex named as $V_k$ center will be formed. From this point, two scintillation mechanisms will proceed. The $V_k$ center can migrate from one site to another in the crystal lattice through thermal activation until it is captured by a $Ce^{3+}$ ion to form $Ce^{4+}$ or a $Ce^{3+}$-$V_k$ associated complex. The free electron is captured subsequently, leading to luminescence. This process is referred to as binary $V_k$-electron diffusion. Before trapped by $Ce^{3+}$, the $V_k$ center may capture an electron from the conduction band to form a STE which itself is luminescent with a decay time on the order of microseconds. Finally, CVL may occur when the incident high-energy photon excites an electron from the upper core band ($3pCl^-$) into the conduction band ($5pCl^-$).

To verify the scintillation mechanisms under neutron and gamma-ray excitation, we used the exponential decay curves to fit the standard pulses (see Fig.7). Each decay time constant of exponential component represents the decay time of different scintillation mechanism. The fitting result of gamma waveform reveals four exponential decay components. Their decay times are 2 ns, 50 ns, 420 ns, and 3400 ns, corresponding to CVL, $Ce^{3+}$, $V_k$, and STE respectively. For neutron waveforms, only $V_k$, and STE are observed: 390 ns and 1500 ns (see Table I).

Table1.　Measured decay times for gamma and neutron induced CLYC emission

| Types | CVL /ns | $Ce^{3+}$ /ns | $V_k$ /ns | STE /ns |
|---|---|---|---|---|
| Gamma-ray | 2 | 50 | 420 | 3400 |
| Neutron | - | - | 390 | 1500 |

## 4  Conclusion

CLYC crystal has excellent properties, and can be a replacement of $^3$He material in homeland security and nuclear safeguards. The capability of robust discrimination between neutrons and gamma-rays makes CLYC a radiation detector with dual neutron and gamma-ray sensitivity[20-21]. Also, based on $^6$Li(n,α) reaction and $^{35}$Cl(n,p) reaction , CLYC can discriminate between thermal neutrons and fast neutrons through pulse height discrimination. In this study, we evaluated the gamma-ray energy resolution and PSD capability of CLYC crystal grown in BGRI. We also achieved the decay times of different scintillation mechanism by fitting decay functions to the neutron and gamma-ray waveform structures. The CLYC crystal has a gamma-ray energy resolution of ~4.5% FWHM @662keV, and has good PSD capability with FoM ~2.6. There is ultrafast scintillation decay (about 2 ns) in CLYC under gamma-ray excitation whereas there is no evidence of ultrafast decay under thermal neutron excitation. The CLYC crystal can be used in neutron-gamma dual mode radiation detector, and can be applied to detect radiation in space. In future work, we plan to grow CLYC crystal doped with enriched $^6$Li, larger size and better properties by optimizing crystal growth technology.

———————————————